\documentclass[12pt]{article}
\usepackage{cite,graphicx,amsmath,amssymb,bbm}

\newcommand{\be}{\begin{equation}}  
\newcommand{\ee}{\end{equation}}  
\newcommand{\bea}{\begin{eqnarray}}  
\newcommand{\eea}{\end{eqnarray}}  
\newcommand{\ol}[1]{\overline{#1}}
\newcommand{\hc}{+\,\mathrm{h.c.}}
\newcommand{\vev}[1]{\langle #1 \rangle}

\newcommand{\tr}{\operatorname{tr}}

\begin{document}

\thispagestyle{empty}
\vspace*{.2cm}
\noindent
HD-THEP-07-11 \hfill 15 May 2007

\vspace*{1.5cm}

\begin{center}
{\Large\bf A natural renormalizable model 
\\[.5cm]
of metastable SUSY breaking}
\\[2cm]
{\large Felix Br\"ummer}\\[1cm]
{\it Institut f\"ur Theoretische Physik, Universit\"at Heidelberg,
Philosophenweg 16 und 19, D-69120 Heidelberg, Germany}
\\[.5cm]
{\small\tt (\,f.bruemmer@thphys.uni-heidelberg.de)}
\\[2.0cm]

{\bf Abstract}
\end{center} 

\noindent 
We propose a model of metastable dynamical supersymmetry breaking in which all scales are generated dynamically. Our construction is a simple variant of the Intriligator-Seiberg-Shih model, with quark masses induced by renormalizable couplings to an auxiliary supersymmetric QCD sector. Since all scales arise from dimensional transmutation, the model has no fundamental dimensionful parameters. It also does not rely on higher-dimensional operators.

\newpage

\section{Introduction}

Recently the idea of metastable dynamical supersymmetry breaking has been revived, starting with the work of Intriligator, Seiberg and Shih (ISS) \cite{Intriligator:2006dd}. ISS showed that fairly simple theories can have metastable vacua with dynamically broken SUSY. Their prototype example is SU($N_c$) SQCD with $N_f$ massive quark flavours, where \mbox{$3N_c/2 > N_f > N_c$} and the quark masses are much smaller than the strong-coupling scale $\Lambda$ of the gauge theory. As already mentioned in \cite{Intriligator:2006dd} and proposed in a broader context in \cite{Dine:2006gm}, such a small mass parameter in a model of metastable SUSY breaking can be dynamically generated by coupling the theory to an additional gauge sector via higher-dimensional operators. For the ISS model (or rather a modified version \cite{Murayama:2006yf} including also gauge-mediation messenger fields) this mechanism was worked out in \cite{Aharony:2006my}: Denoting the field strength superfield of the auxiliary sector by $W_\alpha'$, a coupling of the quarks and antiquarks $q,\tilde q$ of the form
\be
{\cal L}\supset \int d^2\theta\;\frac{\tr q\tilde q}{M^2} \tr W'_\alpha {W'}^\alpha\hc
\ee
leads to quark masses $m\sim {\Lambda'}^3/M^2$ after gaugino condensation. Here $\Lambda'$ is the strong-coupling scale of the auxiliary gauge theory, and $M$ is a high scale at which the theory must be UV-completed, e.g.~the Planck scale if one imagines the model to be embedded in a theory of quantum gravity. In this model, $m\ll\Lambda$ can be easily accomplished, and thus the ISS analysis applies. Several particle physics models with metastable SUSY breaking (see, for instance, \cite{Banks:2006ma}) use similar mechanisms for generating small scales dynamically from higher-dimensional operators. 

In this paper we propose to go one step further, by constructing a model which does not rely on unknown physics at some UV completion scale. We generate a small ISS quark mass scale dynamically from the coupling to an auxiliary sector, but using only renormalizable operators. Our model does not have any dimensionful parameters --- all scales are generated by dimensional transmutation. It consists of two SQCD sectors, the ISS sector and the auxiliary sector, with their matter fields coupled to an additional singlet $S$. $S$ obtains a vacuum expectation value from strong gauge dynamics in the auxiliary sector, which generates an effective mass term for the ISS sector quarks from a superpotential term
\be
W\supset \lambda S\tr q\tilde q.
\ee
To obtain sufficiently small quark masses $m\ll\Lambda$, the dimensionless coupling $\lambda$ must be taken to be moderately small (for the metastable vacuum to survive much longer than the age of the universe, $\lambda\sim 10^{-2}$ is sufficient in a realistic setup). However, we stress that this tuning is rather mild and concerns a dimensionless coupling only. The hierarchy between the fundamental scale and the SUSY breaking scale is still mainly generated by nonperturbative gauge dynamics, which after all is the central idea behind dynamical SUSY breaking.

This paper is organized as follows: In Section 2, we give a brief review of metastable SUSY breaking within the ISS model. In Section 3, we show how small quark masses can be generated by coupling the ISS model to an auxiliary sector. We also give an explicit example to show how this scenario can be realized with plausible choices of parameters. We conclude in Section 4. In the appendix we show that the superpotential of our auxiliary sector is constrained by the symmetries and holomorphy to take the form we have assumed.

\section{The ISS model}
We will now briefly review the analysis of ISS \cite{Intriligator:2006dd}. Consider $N=1$ rigidly supersymmetric QCD with $N_c$ colours and $N_f$ flavours of massive quarks and antiquarks $q^i,\tilde q_i$ ($i=1\ldots N_f$), where $3 N_c/2 > N_f > N_c$. Let us take the quark masses to be equal for simplicity and denote them by $m$. Assume also that $m\ll\Lambda$, where $\Lambda$ is the strong-coupling scale of the gauge theory. The theory is asymptotically free. It has a dual description \cite{Seiberg:1994pq} on scales much lower than $\Lambda$ in terms of an IR free $SU(N_f-N_c)$ gauge theory with $N_f$ dual quarks and antiquarks $\varphi^i,\tilde\varphi_i$ and $N_f^2$ uncharged mesons $\Phi^i_j$. In the dual theory, near the origin of field space the K\"ahler potential is smooth and hence can be taken to be canonical to leading order (up to normalization factors of order one, which we drop). The infrared superpotential is, up to ${\cal O}(1)$ coefficients,
\be\label{IRW}
W=\tilde\varphi^c_i\Phi^i_j \varphi_c^j-m\Lambda \Phi^i_i+\left(\frac{\det \Phi}{\Lambda^{3N_c-2N_f}}\right)^{\frac{1}{N_f-N_c}}\qquad(c=1\ldots N_f-N_c,\:\;i,j=1\ldots N_f).
\ee
At small field values, we can neglect the last term in $W$ because of the $\Lambda$-suppression; then the $F$-terms of $\Phi$ are\footnote{By a common abuse of notation, we use the same symbols for the lowest components of chiral superfields as for the respective superfields themselves.} 
\be 
F_{\Phi^i_j}=\tilde\varphi^c_i \varphi_c^j-m\Lambda\delta_i^j.
\ee
They cannot all vanish because $\tilde\varphi^c_i \varphi_c^j$ has rank $N_f-N_c$, whereas $\delta_i^j$ has rank $N_f$. It turns out that there is a SUSY breaking local minimum, the ISS vacuum, at
\be
\Phi=0,\qquad\qquad (\tilde\varphi^c_i)=(\varphi_c^j)^T=\left(\begin{array}{c} m\mathbbm{1}_{N_f-N_c} \\ 0 \end{array}\right).
\ee
Here $\mathbbm{1}_{N_f-N_c}$ denotes the ${(N_f-N_c)\times(N_f-N_c)}$ unit matrix. At tree-level, the potential still has several flat directions. Those that correspond to Goldstone directions from spontaneously broken global symmetries are unaffected by quantum corrections. The others are lifted by the one-loop Coleman-Weinberg potential, such that the ISS vacuum is indeed locally stable. In addition to the ISS vacuum there are supersymmetric vacua, which are found by taking into account also the determinant term in \eqref{IRW}. However, they are well separated in field space from the ISS vacuum if $m/\Lambda$ is sufficiently small, hence the ISS vacuum can be very long-lived. More precisely, in \cite{Intriligator:2006dd} the bounce action for overcoming the tunneling barrier and decaying into the proper vacuum was estimated to be
\be
S_{\rm bounce}\approx \left(\frac{\Lambda}{m}\right)^{\frac{6 N_c-4 N_f}{N_c}}, 
\ee
which shows that for $m\ll\Lambda$ the lifetime of the ISS vacuum is parametrically large.

\section{Generating small quark masses}
Let us first describe what will eventually become the auxiliary sector of our model. Take SU($N_c'$) SQCD with $N_f'$ flavours of massless quarks and antiquarks $Q,\tilde Q$, where $N_c'>N_f'$. Couple this theory to an additional singlet $S$ with tree-level superpotential
\be
W_{\rm tree}=\lambda'S\tr Q\tilde Q-\kappa S^3.
\ee
In the quantum theory, an additional contribution to the superpotential is generated nonperturbatively \cite{Affleck:1983mk}, which becomes relevant in the infrared:
\be
W_{\rm np}=a\left(\frac{{\Lambda'}^{3N_c'-N_f'}}{\det Q\tilde Q}\right)^{\frac{1}{N_c'-N_f'}}.
\ee
Here $\Lambda'$ is the strong-coupling scale of the gauge theory, and $a$ is a renormalization-scheme dependent number of order one. In \cite{Seiberg:1993vc} it was shown that by holomorphy and symmetry the exact low-energy effective superpotential is $W=W_{\rm tree}+W_{\rm np}$, in a range of parameters where $S$ is the only light degree of freedom and the quarks are integrated out. In the appendix, we show that $W=W_{\rm tree}+W_{\rm np}$ is indeed exact even in the general case. 

To analyze the IR behaviour of the theory, we introduce the meson fields
\be
M^i_j=\frac{1}{\Lambda'}Q^i\tilde Q_j
\ee
(with a trace over colour indices implied), normalized by the $1/\Lambda'$ factor to have canonical dimension. In terms of the mesons and the singlet, the exact low-energy effective superpotential is then
\be\label{Wex}
W_{\rm eff}=\lambda'\Lambda' S\tr M-\kappa S^3+a\left(\frac{{\Lambda'}^{3N_c'-2N_f'}}{\det M}\right)^{\frac{1}{N_c'-N_f'}}.
\ee
%We again assume a canonical K\"ahler potential, up to ${\cal O}(1)$ normalization factors. 
The equations for supersymmetric vacua,
\be
\lambda'\Lambda'\tr M-3\kappa S^2=0,\qquad\lambda'\Lambda' S\,\delta^i_j-\frac{a}{N_c'-N_f'}\left(\frac{{\Lambda'}^{3N_c'-2N_f'}}{\det M}\right)^{\frac{1}{N_c'-N_f'}}\left(M^{-1}\right)^i_j=0,
\ee
are solved by
\be\label{susyvac}
\begin{array}{l}S=b\Lambda'\,e^{\frac{2\pi in}{3N_c'-N_f'}},\\
M=c\Lambda'\,e^{\frac{4\pi i n}{3N_c'-N_f'}}\,\mathbbm{1}_{N_f}, \end{array}
\qquad\qquad (0\leq n< 3N_c'-N_f'),
\ee
where $b$ and $c$ are numerical constants given by
\be
b=\left[\left(\frac{N_f'}{3\kappa}\right)^{N_c'}(\lambda')^{N_f'}\left(\frac{a}{N_c'-N_f'}\right)^{N_c'-N_f'}\right]^{\frac{1}{3N_c'-N_f'}},\quad c=\left[\frac{3\kappa}{(\lambda')^3 N_f'}\left(\frac{a}{N_c'-N_f'}\right)^2\right]^{\frac{N_c'-N_f'}{3N_c'-N_f'}}.
\ee
For simplicity, in the following we choose the couplings $\lambda'$ and $\kappa$ such that $b=c=1$.

We now couple this model to an ISS sector, with the ISS quark mass coming from the expectation value of $S$. The combined superpotential in the UV is
\be\label{WUV}
W=-\lambda\,S\tr q\tilde q+\lambda'\,S\tr Q\tilde Q-\kappa S^3.
\ee
We have deliberately omitted all possible operators with dimensionful couplings here: No scales are introduced by hand. The absence of linear and quadratic terms in $W$ can be further justified by imposing an obvious discrete $\mathbb{Z}_3$ symmetry acting on the chiral superfields, which will be spontaneously broken by nonperturbative effects. 

Let us assume that $\lambda\ll 1$, such that also $\lambda\Lambda\ll \Lambda'$ and $\lambda\Lambda'\ll\Lambda$ (this can of course be achieved by, for instance, choosing the numbers of colours and flavours and the gauge couplings at the renormalization scale such that $\Lambda\approx\Lambda'$, and then setting $\lambda\ll 1$). 

The resulting model has various effective descriptions at different energy scales. In the far UV the appropriate superpotential is \eqref{WUV}. The ISS and auxiliary sector then have effective descriptions at scales below their respective strong coupling scales $\Lambda$ and $\Lambda'$ (either of which can be the higher one): At scales around $\Lambda$ we should pass to the Seiberg dual of the $q$ sector, replacing 
\be\label{dualW}
-\lambda S\tr q\tilde q\;\rightarrow\; -\lambda \Lambda S\tr\Phi+\tr\tilde\varphi\Phi\varphi+\left(\frac{\det \Phi}{\Lambda^{3N_c-2N_f}}\right)^{\frac{1}{N_f-N_c}}.
\ee
Here we anticipate that $S$, which is a dynamical field up to now, will eventually acquire an expectation value, such that the $\lambda S\tr q\tilde q$ term will become an ISS quark mass term.
At scales below $\Lambda'$ the $Q$ sector together with $S$ can be described by the exact superpotential \eqref{Wex}, with the coupling to the $q$ sector viewed as a small perturbation. We should therefore replace
\be
\lambda'\,S\tr Q\tilde Q-\kappa S^3\;\rightarrow\;\lambda'\Lambda' S\tr M-\kappa S^3+a\left(\frac{{\Lambda'}^{3N_c'-2N_f'}}{\det M}\right)^{\frac{1}{N_c'-N_f'}}.
\ee
At scales much below $\Lambda'$, $M$ and $S$ are massive and should be integrated out. Taking for definiteness the phases in \eqref{susyvac} to vanish, we obtain
\be
\vev{S}=\Lambda'\,\left[1+{\cal O}\left(\frac{\lambda^2\Lambda^2}{(\Lambda')^2}\right)\right].
\ee
The correction terms of higher order in $\lambda\Lambda/\Lambda'$ are small by assumption. 

In the IR, the only light degrees of freedom remaining are now the ISS mesons and dual quarks, whose interactions at low energies are governed by the superpotential (dropping again, as in Section 2, the irrelevant last term in \eqref{dualW})
\be
W=-\lambda\vev{S}\Lambda\tr\Phi+\tr\tilde\varphi\Phi\varphi.
\ee
This is just the infrared superpotential of the ISS model from Section 2 with quark mass $m=\lambda\Lambda'+{\cal O}\left(\lambda^3\Lambda^2/\Lambda'\right)$, which is much smaller than $\Lambda$ as required.

Let us illustrate this discussion with a numerical example: Take $N_c=5$, $N_f=6$, $N_c'=4$, $N_f'=3$. Choose the gauge couplings at the Planck scale $M_P= 10^{19}$ GeV as
\be
\alpha(M_P)\equiv\frac{g^2(M_P)}{4\pi}=\frac{1}{42},\qquad\qquad\alpha'(M_P)\equiv\frac{{g'}^2(M_P)}{4\pi}=\frac{1}{45}.
\ee
This gives $\Lambda\approx 1.8\cdot 10^6$ GeV and $\Lambda'\approx 2.3\cdot 10^5$ GeV.
Choosing $\lambda=10^{-2}$, we have $\lambda\Lambda/\Lambda'\approx 8\cdot 10^{-2}$ and $m/\Lambda=\lambda\Lambda'/\Lambda\approx 10^{-3}$, so both these parameters are indeed small. A very crude estimate of the lifetime of the vacuum can be done with the bounce action
\be
S_{\rm bounce}\approx \left(\frac{\Lambda}{m}\right)^{\frac{6}{5}}\approx 3\cdot 10^3.
\ee
With the decay width per unit volume suppressed as
\be
\frac{\Gamma}{V}\frac{1}{m^4}\sim e^{-S_{\rm bounce}},
\ee
the minimal bounce action for our universe to survive for $\approx 10^{10}$ years in a metastable state is only roughly $S_{\min}\approx 400$, so our vacuum is sufficiently long-lived.

The SUSY breaking scale is at about $6\cdot 10^4$ GeV, of the right order of magnitude to be compatible with gauge mediation. Indeed it should be possible to couple our model to a messenger sector to construct a simple gauge-mediated model along the lines of \cite{Murayama:2006yf}.

\section{Conclusions}
We have presented a mechanism by which the ISS model of metastable dynamical SUSY breaking can be made fully natural. In the original ISS model the required small mass scale was put in by hand, and in subsequent refinements generated from higher-dimensional operators, relying on some unknown physics at the UV-completion scale of the theory. Here we have generated the small mass scale from strong gauge dynamics in an auxiliary sector, coupled to the ISS model by renormalizable operators which involve an additional singlet.  A parameter $\lambda$ is required to be moderately small (we have seen that $\lambda\approx 10^{-2}$ is acceptable in an example). 

A possible direction for further work would be to employ this mechanism in a realistic model of particle physics, involving also messenger and visible sector fields. Furthermore, it would be interesting to find a stringy realization of the model presented here, e.g.~arising from an intersecting brane model in type IIA or from branes at singularities in type IIB (see, for instance, \cite{Franco:2006es} for some D-brane constructions of ISS-like models).

\vspace{0.5cm}
\noindent
{\bf Acknowledgements}:\hspace*{.5cm} I would like to thank Stefan Groot Nibbelink, Christoph L\"udeling and especially Arthur Hebecker for useful discussions. I am also obliged to Thomas Dent and Arthur Hebecker for many helpful comments on the manuscript.

\appendix
\section*{Appendix: The exact superpotential for SQCD coupled to a singlet}
Here we derive the exact superpotential of the auxiliary sector as introduced in Section 3.\footnote{Note that our notation here deviates slightly from that used in the main text (we omit the primes and differ by a factor of $\Lambda$ in the definition of the meson field $M$).} Consider SU($N_c$) SQCD with $N_f$ flavours of massless quarks and antiquarks $Q,\tilde Q$, where $N_c>N_f>1$. Let us write down a superpotential which couples the quarks to a singlet field $S$:
\be
W_{\rm tree}=S\lambda\tr Q\tilde Q+\kappa S^3.
\ee
At low energies the theory should be described in terms of the gauge-invariant composites $M^i_j=Q^i\tilde Q_j$. For $\lambda=\kappa=0$, their dynamics is governed by the non-perturbative Affleck-Dine-Seiberg superpotential \cite{Affleck:1983mk}:
\be
W_{\rm np}=a\left(\frac{{\Lambda}^{3N_c-N_f}}{\det M}\right)^{\frac{1}{N_c-N_f}},
\ee
with some scheme-dependent prefactor $a$. We will now argue that the full superpotential, including all quantum corrections, is dictated by symmetry and holomorphy to be
\be\label{exactw}
W=W_{\rm tree}+W_{\rm np}.
\ee
Our line of reasoning is a straightforward variation of the arguments of \cite{Seiberg:1993vc, Intriligator:1994jr}. An exact superpotential was already found in \cite{Seiberg:1993vc} for the system under consideration here, but in a range of parameters where $M$ is heavy enough to be integrated out, so that $S$ is the only relevant low-energy degree of freedom.

At $\kappa=\lambda=0$, the classical theory is invariant under a large symmetry $G$:
\be
G={\rm SU}(N_f)_L\times{\rm SU}(N_f)_R\times{\rm U}(1)_A\times{\rm U}(1)_V\times{\rm U}(1)_R\times{\rm U}(1)_S.
\ee
U$(1)_S$ acts only on the singlet. The other abelian symmetries are an anomalous axial U$(1)_A$, a vectorial $U(1)_V$ and an anomaly-free U$(1)_R$ R-symmetry. The ``baryon number'' $U(1)_V$ will play no part in the following discussion, since all the fields involved at low energies are neutral. (In fact, there are even more symmetries, acting on $S$, which we have already omitted because they will not be relevant.)

We now promote the couplings $\kappa$ and $\lambda$ to classical background chiral superfields, whose nonzero values break $G$. Also, the scale $\Lambda$ of the gauge theory is assigned a charge under the anomalous U$(1)_A$. To be able to write down the most general invariant superpotential, we should promote $\lambda$ to an $N_f\times N_f$ matrix, replacing $\lambda\tr M\,\rightarrow\,\tr(\lambda M)$. The following table lists the charges and dimensions of the fields, couplings and of $\Lambda$, as well as of the basic holomorphic SU$(N_f)_L\times\,$SU$(N_f)_R$-invariants that can be constructed from $M$ and $\lambda$:

\begin{tabular}{c|c|c|c|c|c|c}
& SU$(N_f)_L$ & SU$(N_f)_R$ & U$(1)_A$ & U$(1)_R$ & U$(1)_S$ & dimension \\
\hline
S & $\mathbf{1}$ & $\mathbf{1}$ & 0 & 0 &1 & 1 \\
M & $\mathbf{N_f}$ & $\ol{\mathbf{N}}_{\mathbf{f}}$ & $2$ & $2\frac{N_f-N_c}{N_f}$ & 0 & 2 \\
\hline
$\kappa$ & $\mathbf{1}$ & $\mathbf{1}$ & 0 & 2 & $-3$ & 0 \\
$\lambda$ & $\ol{\mathbf{N}}_{\mathbf{f}}$ & $\mathbf{N_f}$ & $-2$ & $2\frac{N_c}{N_f}$ & $-1$ & 0 \\
$\Lambda$ & $\mathbf{1}$ & $\mathbf{1}$ & $\frac{2 N_f}{3N_c-N_f}$ & 0 & 0 & 1 \\
\hline
$\det M$ & $\mathbf{1}$ & $\mathbf{1}$ & $2 N_f$ & $2N_f-2N_c$ & 0 & $2N_f$ \\
$\det \lambda$ & $\mathbf{1}$ & $\mathbf{1}$ & $-2 N_f$ & $2N_c$ & $-N_f$ & 0 \\
$\tr\left[(\lambda M)^n\right]$ & $\mathbf{1}$ & $\mathbf{1}$ & 0 & $2n$ & $-n$ & $2n$
\end{tabular}

The full superpotential must be holomorphic in both the couplings and the fields. It must have dimension $3$, transform under U$(1)_R$ with charge $2$, and be invariant under the rest of $G$. The most general such function can be written as
\be
W=S\tr\lambda M\:f(I_\alpha),
\ee
where $f$ is a holomorphic function of dimensionless $G$-invariant variables $I_\alpha$. This expression makes sense since the number of independent $I_\alpha$ is finite. In fact, the operators $\tr\left[(\lambda M)^n\right]$ for $n>N_f$ are algebraically dependent on the operators $\tr\left[(\lambda M)^n\right]$ for $n\leq N_f$ (which can be seen e.g.~from Newton's identities). The same is also true for $\det(\lambda M)$. From the table above there are therefore $N_f+4$ independent SU$(N_f)_L\times$SU$(N_f)_R$-invariants, subject to $3$ independent constraints from $U(1)_A\times U(1)_R\times U(1)_S$-invariance and dimensionlessness. Hence there are $N_f+1$ independent $I_\alpha$. We may choose them to be
\be\begin{split}
&I_{0}=\left(\frac{\Lambda^{3 N_c-N_f}}{\det M}\right)^{\frac{1}{N_c-N_f}}(S\tr\lambda M)^{-1},\\
&I_n=\frac{\kappa^n S^{2n}}{\tr\left[(\lambda M)^n\right]}\qquad\qquad\qquad(n=1\ldots N_f).
\end{split}
\ee
At weak coupling, where $\lambda^i_j\,\rightarrow\,0$, $\kappa\,\rightarrow\,0$ and $\Lambda\,\rightarrow\,0$, the superpotential must asymptote to \eqref{exactw}, which means that
\be\label{f}
f = 1+I_0+I_1.
\ee
But since all values of the $I_\alpha$ can be obtained in this limit, \eqref{f} must already be exact, which proves our assertion.

\end{document}